\documentclass[useAMS,usenatbib]{mn2e}

\usepackage{psfig}

\title[MAXI\,J0556$-$332]{The nature of the X-ray transient MAXI\,J0556$-$332}

\author[R. Cornelisse et~al.]{R. Cornelisse$^{1,2}$\thanks{E-mail:
    corneli@iac.es}, P. D'Avanzo$^3$, S. Campana$^3$, J.
  Casares$^{1,2}$, P.A.  Charles$^{4,5}$, 
\newauthor
G. Israel$^6$, T. Mu\~noz-Darias$^3$, K.  O'Brien$^7$, D. Steeghs$^8$, L.
  Stella$^6$, M.A.P. Torres$^9$
  \\
  $^{1}$ Instituto de Astrofisica de Canarias, Via Lactea, La Laguna
  E-38200, Santa  Cruz de Tenerife, Spain\\
  $^2$ Departamento de Astrofisica, Universidad de La Laguna, E-38205
  La Laguna, Tenerife, Spain\\
  $^3$INAF- Osservatorio Astronomico di Brera, via E. Bianchi 46, 23807
  Merate, Italy\\
  $^4$ South Africa Astronomical Observatory, P.O. Box 9, Observatory
  7935, South Africa\\
  $^5$ School of Physics and Astronomy, University of Southampton,
  Highfield, Southampton SO17 1BJ, UK\\
  $^6$ INAF-Osservatorio Astronomico di Roma, Via Frascati 33, I-00040 
  Monteporzio Catone (Rome), Italy\\
  $^7$ Department of Physics, University of California, Santa Barbara, CA, USA\\
  $^8$ Department of Physics, University of Warwick, Coventry, CV4 7AL, UK\\
  $^9$ SRON, Netherlands Institute for Space Research, Sorbonnelaan 2, 3584 CA, Utrecht, The Netherlands\\ 
}
\begin{document}

\date{Accepted Received ; in original form }

\pagerange{\pageref{firstpage}--\pageref{lastpage}} \pubyear{2011}

\maketitle

\label{firstpage}

\begin{abstract}
  Phase-resolved spectroscopy of the newly discovered X-ray transient
  MAXI\,J0556$-$332 has revealed the presence of narrow emission lines
  in the Bowen region that most likely arise on the surface of the
  mass donor star in this low mass X-ray binary. A period search of
  the radial velocities of these lines provides two candidate orbital
  periods (16.43$\pm$0.12 and 9.754$\pm$0.048 hrs), which differ from
  any potential X-ray periods reported.  Assuming that
  MAXI\,J0556$-$332 is a relatively high inclination system that
  harbors a precessing accretion disk in order to explain its X-ray
  properties, it is only possible to obtain a consistent set of system
  parameters for the longer period. These assumptions imply a mass
  ratio of $q$$\simeq$0.45, a radial velocity semi-amplitude of the
  secondary of $K_2$$\simeq$190 km s$^{-1}$ and a compact object mass
  of the order of the canonical neutron star mass, making a black hole
  nature for MAXI\,J0556$-$332 unlikely. We also report the presence
  of strong N\,III emission lines in the spectrum, thereby inferring a
  high N/O abundance. Finally we note that the strength of all
  emission lines shows a continuing decay over the $\simeq$1 month of
  our observations.
\end{abstract}


\section{Introduction}

Amongst the brightest objects in the X-ray sky are the low mass X-ray
binaries (LMXBs), exotic systems where the primary is a compact object
(either a neutron star or black hole) and the secondary a low-mass
($<$1$M_\odot$) star. Since the secondary is overflowing its
Roche-Lobe, matter is accreted via an accretion disk onto the compact
object, giving rise to the observed X-rays. The majority of these
LMXBs, the so-called X-ray transients, only show sporadic X-ray
activity, but most of the time remain in a state of low-level
activity, referred to as quiescence (see e.g. Psaltis 2006 for an
overview and more detailed references).

Due to reprocessing of the X-rays, mainly in the outer accretion disk,
a LMXB also becomes much brighter in the optical during a transient
outburst (e.g. Charles \& Coe 2006). Unfortunately, this reprocessed
emission completely dominates the optical light, making radial
velocity studies using spectral features from the donor star
impossible in most cases, while during their quiescent state most
transients become too faint for such studies. A powerful alternative
method of investigation was opened when Steeghs \& Casares (2002)
detected narrow emission line components that originated from the
irradiated face of the donor star in Sco\,X-1. These narrow features
were most obvious in the Bowen blend (NIII $\lambda$4634/4640 and CIII
$\lambda$4647/4650), which is the result of UV fluorescence from the
hot inner disk, and gave rise to the first radial velocity curve of
the mass donor in Sco X-1. Thus far, high resolution phase-resolved
spectroscopic studies have revealed these narrow lines in more than a
dozen optically bright LMXBs, including several transients during
their outburst, leading (for most of them) to the first constraint on
the mass of the compact object (see Cornelisse et~al.  2008 for an
overview).

\begin{figure*}
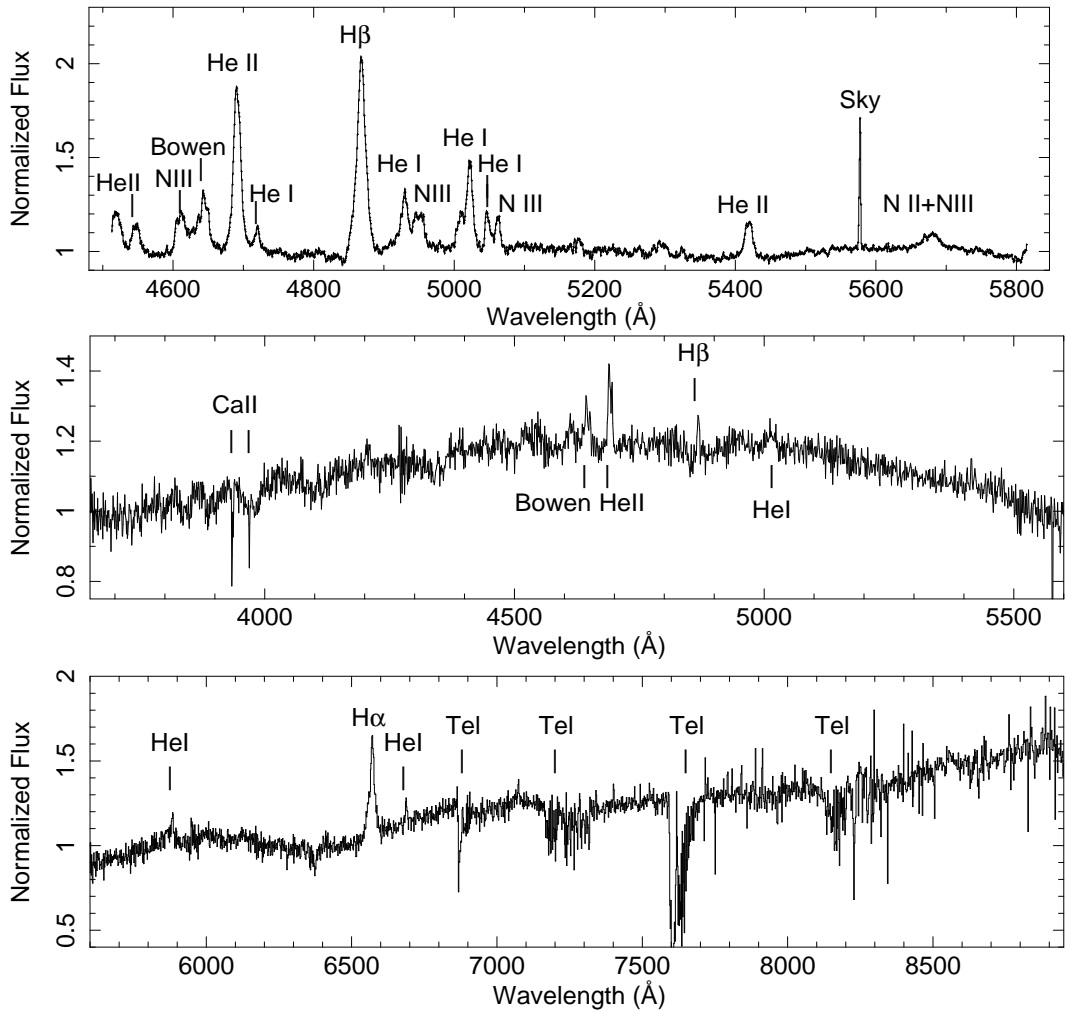
\begin{center}
\psfig{figure=Fors.ps,angle=-90,width=14cm}
\psfig{figure=normalized_ubv.ps,angle=-90,width=14cm}
\psfig{figure=normalized_vis.ps,angle=-90,width=14cm}
\end{center}

\caption{Average flux normalized spectrum of MAXI\,J0556 obtained with
  FORS (top) and X-Shooter (middle and bottom). The X-Shooter spectrum
  is split into two panels, with the top showing the spectrum obtained
  with the ``UBV'' arm the middle panel the ``VIS'' arm. Indicated in
  both spectra are the most prominent emission lines detected, and we
  have also indicated the most prominent Telluric absorption features
  with ``Tel''.
  \label{spec}}
\end{figure*}

On January 11 2011 the X-ray transient MAXI\,J0556$-$332 (hereafter
MAXI\,J0556) was newly discovered by MAXI/GSC (Matsumura et~al. 2011),
and quickly thereafter localized by the Swift X-ray telescope (Kennea
et~al. 2011).  Eclipse-like features in the X-ray light-curves led to
several possible orbital periods (Strohmayer 2011, Maitra
et~al. 2011a). Although follow-up X-ray observations excluded all
except one (9.33 hr) of these periods (Belloni et~al. 2011), further
observations were also able to discard this period (Belloni private
communication). The optical counterpart was quickly confirmed by
Halpern (2011) as an $R$$\simeq$17.8 object, making it an excellent
target for high resolution spectroscopy. Initial studies of the X-ray
spectral and timing variations were unable to indicate the nature of
the compact object in MAXI\,J0556 (Belloni et~al. 2011), but more
detailed studies by Homan et~al. (2011) suggest that its behavior is
similar to some of the neutron stars at high accretion rates (the
so-called Z-sources). Together with its optical and radio properties
this strongly suggests that MAXI\,J0556 harbors a neutron star
(Russell et~al. 2011; Coriat et~al. 2011).

In order to identify the nature of MAXI\,J0556 using the narrow
  components in the Bowen region we triggered our Target
of Opportunity (ToO) observations using FORS 2 on the Very Large
Telescope (VLT) and complemented it with Directors Discretionary Time
(DDT) using X-Shooter, which is also on the VLT. In this paper we
present the results of these observations. In Sect.\,2 we discuss the
observations in detail, and in Sect.\,3 we show the results. We finish
with a discussion in Sect.\,4 and present further evidence that
MAXI\,J0556 is most likely a neutron star LMXB.

\section{Observations and Data Reduction}

\begin{table}\begin{center}
\caption{Observation log of MAXI\,J0556. Indicated are the observing 
dates and times, instrument used (Setting), integration time for each 
spectrum (Int.), number of spectra obtained during each night (No,) and 
the variation of the seeing during these observations (as measured by 
the DIMM on Paranal). 
\label{obs}}
\begin{tabular}{ccllcl}
\hline
Date         & Time & Setting & Int. &  No. & Seeing\\
(yy-mm-dd) & (UT) &         & (s)      &                & (arcsec)\\   
\hline
2011-01-19   & 00:44-04:50 & FORS & 671 & 20 & 0.45-0.97 \\
2011-01-21   & 00:54-01:51 & FORS & 671 & 4 & 0.95-1.14 \\
2011-01-22   & 00:57-01:52 & FORS & 671 & 4 & 0.39-0.45 \\
2011-01-23   & 00:41-01:37 & FORS & 671 & 4 & 0.74-0-98 \\
2011-01-24   & 00:50-01:45 & FORS & 671 & 4 & 0.46-0.62 \\
2011-01-25   & 02:38-03:32 & FORS & 671 & 4 & 0.39-0.77 \\
2011-02-07   & 00:48-02:19 & FORS & 671 & 7 & 0.47-0.62 \\
2011-02-14   & 00:28-01:22 & X-Shooter & 595 & 4 & 1.34-1.69\\
2011-02-14   & 03:45-04:39 & X-Shooter & 595 & 4 & 0.79-1.02\\  
\hline
\end{tabular}
\end{center}\end{table}

On 20 January 2011 we triggered our ToO to observe MAXI\,J0556 over a
period of $\simeq$3 weeks using the FORS 2 instrument on the
ESO/VLT. For each exposure we used the 1400V volume-phased holographic
grism with a slit width of 0.7$''$, leading to a wavelength coverage
of $\lambda$$\lambda$4512-5815 with a resolution of 87 km s$^{-1}$
(FWHM). Furthermore, these observations were complemented with
2$\times$1 hr of X-Shooter DDT observations (also located on the VLT).
Although both blocks were observed on the same night (14 February
2011), they were separated by several hours. For the X-Shooter
observations, which cover the full UV-optical-IR wavelength band, we
used a 1.0$"$ slit for the UBV arm (and 0.9$"$ for the other two
arms), resulting in a resolution of 66 km s$^{-1}$ (FWHM) around the
Bowen region.  In Table\,\ref{obs} we give an overview of the
observations.

For the FORS observations we de-biased and flat-fielded all the images
and used optimal extraction techniques (from the PAMELA software
package) to maximize the signal-to-noise ratio of the extracted
spectra (Horne 1986). Wavelength calibration was performed using the
daytime He, Ne, Hg and Cd arc lamp exposures. We determined the
pixel-to-wavelength scale using a 4th order polynomial fit to 10
reference lines giving a dispersion of 0.64 \AA\,pixel$^{-1}$ and rms
scatter $<$0.01 \AA. The X-Shooter observations were reduced using the
pipeline v.1.2.2 provided by ESO (see Goldoni et~al. 2006 for more
information), resulting in three 1-dimensional spectra (for each arm
of the instrument) per exposure that were flux calibrated using a flux
standard obtained during the same night. For the corresponding
analysis of the full dataset we used the MOLLY package.

 Finally, we reduced our $B$-band acquisition images from the FORS
 observations using standard reduction techniques (i.e. performed bias
 subtraction and flat-fielding). For each image the exposure time was
 5 seconds. We only had one image at the start of each observing
 block, resulting in a total of 7 images. From the reduced
 images we performed a photometric calibration against the USNO
 $B$-band measurements from nearby stars, and list the results in
 Table\,\ref{ew}. We note that for our X-Shooter observations we only
 have a $g$' broad band image, and have not included its magnitude in
 the Table.

\section{Data Analysis}

\subsection{Spectrum}

First we present the average normalized UV-optical spectrum of
MAXI\,J0556 obtained by X-Shooter (top and middle panels) and FORS 2
(bottom) in Fig.\,\ref{spec}, and have indicated the most prominent
lines.  The IR arm of X-Shooter was completely dominated by noise, and
we therefore did not include this part of the spectrum in
Fig.\,\ref{spec}. This figure emphasizes the remarkable contrast in
emission line strength between the FORS and X-Shooter observations.
In the latter only the most prominent emission lines that are
typically observed in LMXBs (i.e the Balmer and He lines and also the
Bowen blend) are present, but they are much weaker than during the
FORS observations.

\begin{table}\begin{center}
\caption{Overview of the average Equivalent Width (EW) of
  He\,II$\lambda4686$ and H$\beta$, the two most prominent emission
  lines in the FORS spectra, during each observing night. Also listed
  is the corresponding $B$ magnitude from the acquisition images
  obtained at the beginning of each block of FORS observations. Also
  included, as the final entry, are the EW measurements from the
  average X-Shooter spectrum.
\label{ew}}
\begin{tabular}{cccc}
\hline
Date         & EW He\,II  & EW H$\beta$ & Magnitude\\
(yy-mm-dd)   & (\AA)       &  (\AA)     & $B$\\
\hline
2011-01-19   & 4.32$\pm$0.02 & 6.15$\pm$0.02 & 17.1$\pm$0.1\\
2011-01-21   & 3.42$\pm$0.03 & 5.61$\pm$0.03 & 17.0$\pm$0.1\\
2011-01-22   & 2.82$\pm$0.02 & 3.19$\pm$0.02 & 17.1$\pm$0.1\\
2011-01-23   & 2.00$\pm$0.03 & 2.81$\pm$0.03 & 16.9$\pm$0.1\\
2011-01-24   & 1.86$\pm$0.03 & 2.64$\pm$0.03 & 17.0$\pm$0.1\\
2011-01-25   & 1.83$\pm$0.02 & 3.31$\pm$0.03 & 17.0$\pm$0.1\\
2011-02-07   & 1.19$\pm$0.02 & 0.79$\pm$0.02 & 16.4$\pm$0.1\\
\hline
2011-02-14   & 1.04$\pm$0.06 & 0.25$\pm$0.06 & --\\
\hline
\end{tabular}
\end{center}\end{table}

For our first observation, the wealth of emission lines in the average
FORS spectrum shown in Fig.\,\ref{spec}, we start by noting that the
typical emission lines for LMXBs (i.e. H$\beta$, He\,II $\lambda$4686,
Bowen and all He\,I lines) are all present. Furthermore, there are
several other emission lines which are not commonly observed in
LMXBs. Although these uncommon lines become fainter as a function of
time in a similar way to H$\beta$ and He\,II $\lambda$4686 (see
above), they appear to be present in all spectra. We therefore
conclude that they are real and not an artifact of our reduction. In
order to identify these lines, we first moved the average spectrum in
such a way that the central wavelength of several He\,I lines
corresponded to their rest-wavelength. Using an atomic line list by
van Hoof \& Verner (1997) we then noted that most of the uncommon
emission lines correspond to the strongest N\,III or N\,II
transitions or their multiplets.  We therefore tentatively conclude
that all these lines are due to N\,III or N\,II and have indicated
them as such in Fig.\,\ref{spec}.

\begin{figure}
\psfig{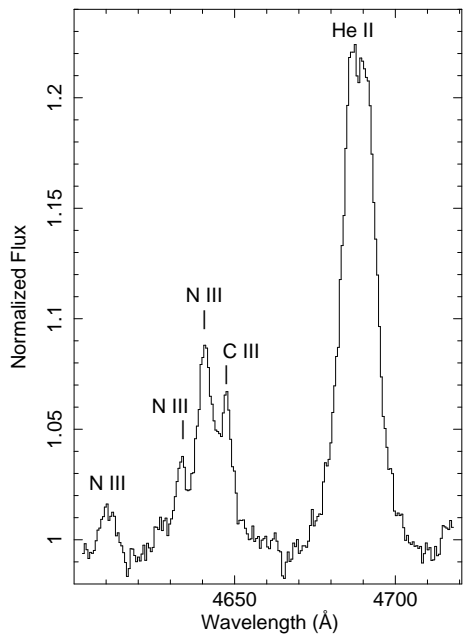}
\caption{Average spectrum of MAXI\,J0556 around the He\,II
  $\lambda$4685.75 and Bowen region, in the rest frame of the donor
  star. The indicated narrow components detected in the Bowen region
  are N\,III $\lambda$4610.55/4610.75 (not visible in the individual
  spectra), N\,III $\lambda$4634.12, N\,III $\lambda$4640.64 and
  C\,III $\lambda$4647.42.
\label{average}}
\end{figure}

Also, given the suggestion by Maitra et~al. (2011b) that MAXI\,J0556
could have an extremely high N/O abundance, we also checked for the
presence of strong O and C lines by cross-correlating the line list
obtained from ultra-compact C/O binaries by Nelemans
et~al. (2004). This shows that only C\,III$\lambda$4652, which is
typically the strongest C or O line in most X-ray binaries (see
e.g. Steeghs \& Casares 2002), is clearly present and its relative
strength is comparable to that of other X-ray binaries.

In order to understand our second observation, the much weaker
emission lines in the X-Shooter, we examined the average FORS spectra
for each individual night in more detail. We note that there is a
large variability in the emission lines from night to night. In order
to quantify this variability we averaged all our spectra during an
observing night together and measured the Equivalent Width (EW) of the
two strongest emission lines (i.e. He\,II$\lambda$4686 and H$\beta$)
and list the result in Table\,\ref{ew}. Interestingly, the EW for both
lines shows a decreasing trend as a function of time (with the
exception of H$\beta$ during January 25), while the $B$-band magnitude
stays more or less constant (except during the last FORS
observation). Although this decrease in EW resembles a power-law
decay, almost all individual points strongly deviate from their best
power-law fit, and we decided against providing any fit results. We
conclude that the strength of the emission lines in MAXI\,J0556 is
independent of the continuum flux, and show a continuing decay over
the $\simeq$1 month of our observations.

\subsection{Radial Velocities}

A close inspection of the Bowen region shows that it consists of
several narrow components, and in Fig.\,\ref{average} we show a
close-up of this region. In most individual spectra we could identify
two (sometimes three) individual components. Following previous
observations of the Bowen region (see e.g. Cornelisse et~al. 2008 for
an overview) we identified the strongest component with N\,III
$\lambda$4640.64 and the second strongest as C\,III
$\lambda$4647.42. In the cases that a third component was present this
was identified as N\,III $\lambda$4634.12, but we could not find
evidence for C\,III $\lambda$4650.2, a line that was clearly detected
in e.g. Sco\,X-1 (Steeghs \& Casares 2002).

The narrow components in the Bowen region showed clear velocity shifts
from night to night. In order to obtain radial velocities from them we
averaged two consecutive FORS spectra together to increase the
signal-to-noise, while for the X-Shooter spectra we averaged all
spectra from the same observing block. This provided a total of 26
spectra from which radial velocities could be measured (note that only
the very last spectrum of FORS was not grouped with any other
spectrum). Following Steeghs \& Casares (2002) we fitted the narrow
components with 3 Gaussians (N\,III$\lambda$4634.12/4640.64 and
C\,III$\lambda$4647.42) under the assumption that they all have a
common radial velocity but an independent strength. Using a least
squares technique we determined the best common radial velocity and
corresponding error.

\begin{figure}
\psfig{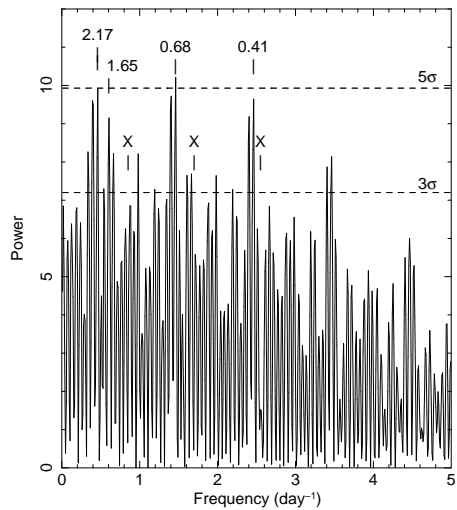}
\caption{Lomb-Scargle periodogram of all our observations of
  MAXI\,J0556.  We have indicated the 3$\sigma$ and 5$\sigma$
  significance levels. Indicated with an X are the proposed X-ray
  periods by Strohmayer (2011) and the corresponding periods (in days)
  of the 4 highest peaks. Note that from our first block of FORS
  observations we can rule out any frequency $\ge$5 cycles per day
  (see Sect.\,3.2).
\label{period}}
\end{figure}

In order to constrain the orbital period we performed a period
analysis on the derived radial velocities using the Lomb-Scargle
method (Scargle 1982), which is best suited for unevenly sampled
time-series. In Fig.\,\ref{period} we show the resulting periodogram
where we have indicated the 4 strongest peaks that represent good
candidates for the orbital period (note that most are related to each
other due to the daily alias).

To estimate the significance of the peaks in the periodogram we
performed a Monte-Carlo simulation, using identical temporal sampling
as for the original radial velocity measurements. Furthermore, our
simulations used a distribution and mean of radial velocities that was
similar to the original dataset. We produced 500,000 random radial
velocity datasets and measured the peak power from the corresponding
periodograms. From the distribution of the peak powers we estimated
both the 3$\sigma$ and 5$\sigma$ confidence levels and have indicated
these in Fig.\,\ref{period}. The only periods that have a significance
$\ge$5$\sigma$ are 2.17 and 0.68 days, and we note that the (almost
significant) 0.41 day period is related via the daily alias to the
other two significant periods.

\begin{figure}\begin{center}
\psfig{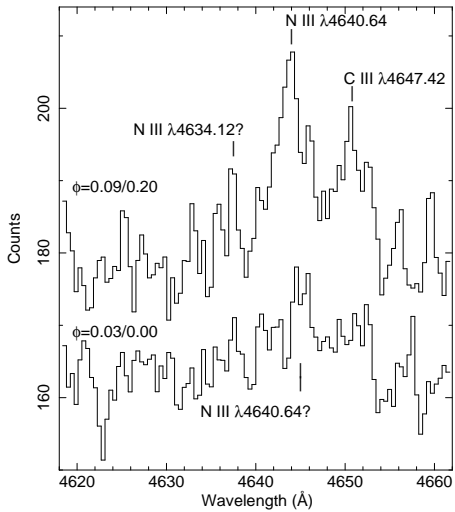}
\end{center}
\caption{Average spectrum of the first (bottom) and second (top)
  X-Shooter observations showing a close-up of the Bowen region.  We
  list the corresponding orbital phases for a 1.65 day (first) and
  0.68 day (second) orbital period for each block, and have indicated
  the most prominent narrow components. Note that the narrow
  components are shifted from their rest-wavelength due to radial
  velocity motion, and the lines indicated with a question mark are
  too noisy for unambiguous identification as a true narrow component.
\label{phase}}
\end{figure}

First of all we note that none of the suggested X-ray periods by
Strohmayer (2011) fit our data.  Furthermore, our first block of 20
FORS spectra consisted of 3.7 hrs of continuous observations, with
radial velocity measurements that only show an increasing trend and a
total amplitude of $\simeq$60 km s$^{-1}$. Since this amplitude is
much smaller than what is observed from night to night ($\simeq$200 km
s$^{-1}$) we conclude that the orbital period must be longer than 0.2
days. Also, the spectra in Fig.\,\ref{spec} show strong Hydrogen
lines, thereby excluding an ultra-compact nature of MAXI\,J0556 and
hence any period $\le$1 hr (i.e. close to the integration time of our
spectra) as was suggested by Maitra et~al. (2011b). We therefore
conclude that one of the periods indicated in Fig.\,\ref{period} is
likely to be the true orbital period.

We also looked in more detail at the two blocks of X-Shooter
observations, and Fig.\,\ref{phase} shows the average spectrum around
the Bowen region of both blocks. We note that the narrow lines, in
particular N\,III $\lambda$4640, are almost absent during the first
block but have become stronger during the second (while the strength
of emission lines such as H$\beta$ and He\,II $\lambda$4686 have not
changed significantly). In order to explain this change in the narrow
components in only a few hours, we estimated (for the 4 potential
orbital periods listed in Fig.\,\ref{period}) the orbital phases for
both blocks of X-Shooter observations. First of all we note that that
for all orbital periods the first block was obtained around orbital
phase 0 (i.e. when the donor star is closest to us). However, the
second block was taken around orbital phase 0.25-0.35 for the shorter
periods (0.68 and 0.41 days), but is still around phase 0 for the
longer periods (2.17 and 1.65 days).

Under the assumption that the narrow components in the Bowen region
arise on the irradiated surface of the donor star and the inclination
is relatively high (see Sect. 4), the simplest way to explain the
change in line strength between the two blocks of X-Shooter
observations is that our view of the irradiated surface has changed.
Since the first block was obtained around orbital phase 0, the
irradiated side should be least visible, and this could explain the
weakness of the narrow lines. To explain the increase in line
strength, the second block must then have been observed at a
significantly different orbital phase where our view of the irradiated
surface must have improved. Since the orbital phase only changes
significantly between the X-Shooter blocks for the shorter orbital
periods (0.68 and 0.41 days), we conclude that the orbital period must
be one of these two. Combined with the fact that the timing of the
eclipse-like features (see Sect. 4 for a discussion on their nature)
also suggests a period $\le$1.2 days, we propose that the true orbital
period is either 0.684$\pm$0.005 or 0.406$\pm$0.002 days.

\begin{figure}
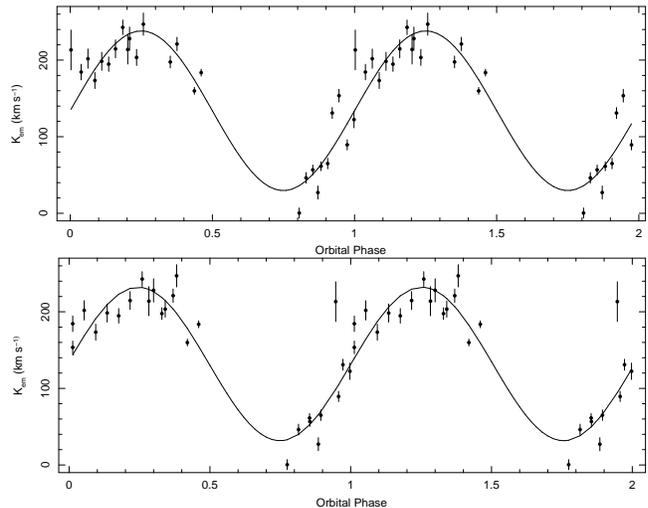

\psfig{figure=rv_68.ps,angle=-90,width=8.5cm}
\psfig{figure=rv_041.ps,angle=-90,width=8.5cm}
\caption{Phase-folded radial velocity curve of MAXI\,J0556 obtained
  from the narrow Bowen components for an orbital period of 0.68d
  (top) and 0.41d (bottom). For clarity, each curve is shown twice.
\label{rv}}
\end{figure}

Unfortunately it is not possible to distinguish between these two
final candidate periods and we assume that both are equally likely. In
Fig.\,\ref{rv} we show the phase-folded radial velocity curve for both
candidate periods, and in Table\,\ref{parm} we list the best fitting
parameters of their radial velocity curves.

\begin{figure}
\psfig{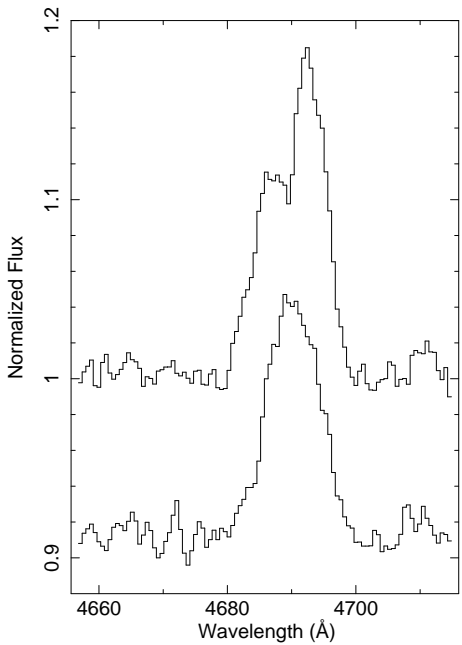}
\caption{Close-up of the average He\,II $\lambda$4686 emission line
  for two different observing nights. Both observations were obtained
  at similar orbital phases (for a 0.68d and 0.41d period), but show
  different line profiles and strengths, Note that for illustration
  purposes we have off-set the spectra from each other.
\label{change}}
\end{figure}

Finally, due to the highly variable nature of the line profiles and
strengths of He\,II $\lambda$4686 and H$\beta$ (see Table\,\ref{ew}
and also illustrated in Fig.\,\ref{change}) we did not attempt to
obtain an estimate of the radial velocity semi-amplitude of the
compact object. However, we do point out that both lines appear to be
red-shifted compared to the narrow lines in the Bowen region. Whereas
N\,III $\lambda$4641 shows a systemic velocity of 130 km s$^{-1}$
(slightly dependent on the assumed period, see Table\,\ref{parm}),
He\,II shows a mean velocity of 340 km s$^{-1}$ and for H$\beta$ it is
already 400 km s$^{-1}$. Whereas for H$\beta$ such behavior has been
observed before (see e.g. Cornelisse et~al. 2007), and could be due to
the presence of other non-resolved emission lines, it is uncommon for
He\,II. Since the lines do not appear to be asymmetric and there is
also no indication for the presence of a P\,Cygni-like profile, we
currently do not have a good explanation for this shift.

\section{Discussion}

We obtained optical spectroscopic observations of MAXI\,J0556 during
its outburst in 2011 and have shown that the average spectrum is not
only dominated by strong H, He emission lines but also N. Although the
presence of Hydrogen does rule out an ultra-compact nature as was
suggested by Maitra et~al. (2011b), it does support their suggestion
that MAXI\,J0556 has an unusually high N/O abundance.

Another interesting result is the power-law decay of all the emission
lines over the time-span of our observations. This decay appears to be
in contradiction with the results by Fender et~al. (2009), which
showed an anti-correlation between the EW of H$\alpha$ and the
continuum emission. However, they were concerned with the fading phase
of the transient outburst, while our observations are mainly during
the peak of the outburst.  It is interesting to note that in Fig.\,2
of Fender et~al. (2009) the H$\alpha$ EW during the peak of the
outburst also appears to be dropping for most transients. Fender
et~al. (2009) provide several explanations, and most of them can also
be adopted here. For example, it could be possible that
the outer accretion disk becomes hotter over time, thereby changing the
optical depth of the emission lines, although saturation effects that
hamper the production of the emission lines or even spectral and
geometrical changes cannot be ruled out.

Furthermore we have detected narrow Bowen lines in the optical
spectrum that moved from night to night. Following other LMXBs for
which similar narrow features have been seen (e.g. Cornelisse
et~al. 2008), we think it likely that these lines arise on the
irradiated surface of the donor star of MAXI\,J0556. Our period
analysis shows that there are two possible orbital periods, 0.68d and
0.41d. From these periods we derived a radial velocity semi-amplitude
of 101-104 km s$^{-1}$.

\begin{table}\begin{center}
\caption{Overview of the system parameters of MAXI\,J0556 for both 
candidate orbital periods. $T_0$ indicates the time of inferior 
conjunction of the donor star.
  \label{parm}}
\begin{tabular}{lll}
\hline
  & 16 hrs & 9.75 hrs\\
\hline
$P_{\rm orb}$ (day)        &  0.684(5) & 0.406(2)\\
$T_0$ (HJD-2,450,000)     &  5,581.520(3) & 5,581.541(2)\\
$K_{\rm em}$ (km s$^{-1}$) & 104.2$\pm$3.8 & 100.8$\pm$3.3\\
$\gamma$  (km s$^{-1}$)   & 133.7$\pm$1.9 & 131.5$\pm$1.8\\
$f$($M_X$) ($M_\odot$)     & $\ge$0.07         & $\ge$0.04\\
\hline
\end{tabular}
\end{center}\end{table}

If the narrow components truly trace the irradiated companion we
thereby obtain a lower limit to the mass-function of
$f$$(M_X)$=$M_X$sin$^3i$/(1+$q$)$^2$$\ge$0.07$M_\odot$ (for $P_{\rm
  orb}$=0.68 day) or $\ge$0.04$M_\odot$ (for $P_{\rm orb}$=0.41 day),
where $i$ is the inclination and $q$(=$M_X$/$M_{\rm donor}$) is the
mass ratio. These mass functions are true lower-limits, since the
lines should trace the radial velocity of the surface and not the
center of mass of the donor star, and its true orbital velocity
($K_2$) should therefore always be higher (Mu\~noz-Darias
et~al. 2005).

One way we could further constrain the system parameters is by taking
account of the X-ray properties of MAXI\,J0556, and in particular the
reported eclipse-like features (Strohmayer 2011; Maitra
et~al. 2011a). Strohmayer (2011) reported that the eclipse was not
total, nor ``sharp'', as might be expected if the X-ray source is
obscured by the donor star. Belloni (private communication) confirmed
this and suggested that these features more resemble dips, as are also
observed in the high inclination system 4U\,1254$-$69 (e.g. Diaz-Trigo
2009). In 4U\,1254$-$69 these dips are thought to be caused by
obscuration of the central X-ray source by the outer accretion disk
edge, and the depth and strength strongly varies (sometimes the dips
are even completely absent). Diaz-Trigo (2009) suggested the presence
of a tilted/precessing accretion disk in 4U\,1254$-$69 and if
something similar is present in MAXI\,J0556 it implies a relatively
high inclination.

The presence of such a tilted/precessing accretion disk would make
MAXI\,J0556 similar to the dwarf novae that show superhumps, and
could also explain why neither of our two potential orbital periods is
close to any period suggested by the X-ray dips (Strohmayer
2011). Superhumps are commonly observed in dwarf novae during a
superoutburst (see e.g.  Patterson et~al. 2005 for an overview), and
are thought to be caused by an instability at the 3:1
resonance that forces the eccentric accretion disk to precess (e.g.
Whitehurst 1988). Due to these deformations of the disk shape the
photometric period in dwarf novae is typically a few percent longer
than the orbital period, although superhump periods shorter than the
orbital period have also been observed if the precessing disk is
counter-rotating (Patterson et~al. 2002).

Assuming that MAXI\,J0556 is a high inclination system (but not high
enough for the donor to obscure the X-ray source) with a precessing
accretion disk that sometimes partly obscures the central X-ray
source, we can try to further constrain its system parameters.  First
of all, one of the original reported periods by Strohmayer (2011) must
then be the superhump period. If the orbital period is 0.68 days (16
hrs) the superhump period would most likely be 0.58 days. Following
Patterson et~al.  (2005), this would lead to an observed fractional
period excess of $\epsilon$=0.14 and suggests a mass ratio of
$q$$\simeq$0.45. We note that this mass ratio is high for systems that
typically show precession. However, Osaki (2005) pointed out that
sufficiently hot accretion disks may expand beyond the 3:1 resonance
radius and start precessing. Since MAXI\,J0556 is thought to be a
Z-source and should have accretion rates close to Eddington (Homan
et~al. 2011), therefore it might be reasonable to expect a such hot
accretion disk and therefore precession.

Using $q$=0.45 we can correct the velocity of the irradiated surface
(our observed $K_{\rm em}$ of 104 km s$^{-1}$) to the center of mass
velocity of the donor star using the $K$-correction developed by
Mu\~noz-Darias et~al.  (2005). This would lead to a maximum $K_2$
velocity of 190 km s$^{-1}$ (for a disk with an opening angle of
0$^\circ$). Furthermore, $q$=0.45 suggests that eclipses by the
secondary will occur for an inclination $\ge$72$^{\circ}$ (Paczynski
1974), and we use this as a first estimate of the
inclination. Combining all these system parameters we obtain a mass
for the compact object of $\simeq$1.2$M_\odot$, which is close to the
canonical 1.4$M_\odot$ neutron star mass. We therefore believe that
for a disk opening angle $\simeq$6$^{\circ}$ and a relatively high
inclination it is also possible to obtain a 1.2-1.4$M_\odot$ compact
object.

Given the system parameters above, we would need an inclination of
$\simeq$40$^{\circ}$ to obtain a black hole with a mass of
3.2$M_\odot$. This is too low to realistically expect any dipping
behavior. 

For an orbital period of 0.4 days (with a superhump period of 9.33
hrs), and making the same assumptions as above, we would obtain a mass
for the compact object of $\simeq$0.2$M_\odot$. Again, this would lead
to unrealistically low inclinations to obtain any sensible compact
object mass. We therefore think that our data are most consistent with
a neutron star LMXB in a 0.68 days orbit and being observed at
moderately high inclination.

\section{Conclusions}

We have presented the results of a spectroscopic campaign on the X-ray
transient MAXI\,J0556 close to peak of outburst and have shown
that strong N\,III emission lines are present in the spectrum, while C
and O show no enhancement. This is in agreement with Maitra
et~al. (2011b) who have suggested an unusually high N/O
abundance. Furthermore, all emission lines show a continuing decay
over the $\simeq$one month of our observations, for which we have no
good explanation.

Our radial velocity study has shown that only two orbital periods
(0.68 and 0.41 days) are possible. From our dataset only we cannot
distinguish the true orbital period so both periods are equally
likely. We suggest that MAXI\,J0556 harbors a precessing accretion
disk to explain not only the disappearance of the dip-like features
observed in X-rays, but also the discrepancy between the periods
suggested by Strohmayer (2011) and ourselves. If the presence of a
precessing accretion disk can be proved, then the X-ray dips suggest
not only a reasonably high inclination but also that the longer period
is most likely the true orbital period. This would imply that the
compact object is a neutron star, strengthening the suggestion of
Homan et~al. (2011).
 
Since a superhump is usually not observed when a system is in
quiescence, the scenario outlined above can only be confirmed with
photometric observations obtained during a future outburst. Finding the
true orbital period on the other hand can be obtained by observing
MAXI\,J0556 when it is back in quiescence. With a quiescent optical
magnitude of $R$$\simeq$20, it is easily accessible for both
photometric and spectroscopic studies with medium-sized
telescopes. Such studies would not only constrain the true orbital
period and the radial velocity semi-amplitude of the center of mass of
the donor star (instead of its irradiated surface), but hopefully
constrain the inclination and binary mass ratio. This would offer
solid dynamical constraints on the masses of the binary components, in
order to verify that MAXI\,J0556 indeed harbors a neutron star and
more accurately determine its mass.

\section*{Acknowledgments}
This work is based on data collected at the European Southern
Observatory Paranal, Chile [Obs.Ids. 286.D-5037(A) and
  086.D-0318(A)]. We cordially thank the ESO director for granting
Director's Discretionary Time. RC wants to thank Tomaso Belloni for
providing important information on the X-ray properties of
MAXI\,J0556. We acknowledge the use of PAMELA and MOLLY which were
developed by T.R.  Marsh, and the use of the on-line atomic line list
at http://www.pa.uky.edu/$\sim$peter/atomic.  RC acknowledges a Ramon
y Cajal fellowship (RYC-2007-01046) and a Marie Curie European
Reintegration Grant (PERG04-GA-2008-239142). RC and JC acknowledge
support by the Spanish Ministry of Science and Innovation (MICINN)
under the grant AYA 2010-18080. This program is also partially funded
by the Spanish MICINN under the consolider-ingenio 2010 program grant
CSD 2006-00070. TMD acknowledges funding from the European Community's
Seventh Framework Programme (FP7/2007-2013) under grant agreement
number ITN 215212. DS acknowledges an STFC Advanced Fellowship.

\bsp

\label{lastpage}

\end{document}